

\input harvmac

\overfullrule=0pt


\def\C{{\scriptscriptstyle C}}

\def\J{{\scriptscriptstyle J}}

\def\P{{\scriptscriptstyle P}}
\def\Q{{\scriptscriptstyle Q}}



\def\a{\alpha}
\def\b{\beta}
\def\c{\gamma}
\def\d{\delta}
\def\e{\epsilon}
\def\g{\gamma}

\def\s{\sigma}
\def\t{\tau}
\def\th{\theta}
\def\u{\mu}
\def\v{\nu}


\def\aS{\alpha_s}

\def\Br{{\rm Br}}
\def\cbar{{\overline c}}
\def\ccdot{\hbox{\kern-.1em$\cdot$\kern-.1em}}
\def\Dbar{{\overline D}}

\def\GeV{\>\, \rm GeV}

\def\gtap{\raise.3ex\hbox{$>$\kern-.75em\lower1ex\hbox{$\sim$}}}
\def\J{J/\Psi}

\def\ltap{\raise.3ex\hbox{$<$\kern-.75em\lower1ex\hbox{$\sim$}}}

\def\Meta{M_{\eta_c}}
\def\MeV{\> {\rm MeV}}
\def\mQ{m_\Q}

\def\Nc{N_c}
\def\nf{n_f}

\def\onedtwo{{^1\hskip-0.2 em D_2}}
\def\onepone{{^1\hskip-0.2 em P_1}}
\def\oneszero{{^1\hskip-0.2 em S_0^{(1,8)}}}
\def\pbar{{\overline{p}}}

\def\pperp{p_\perp}

\def\Qbar{{\overline Q}}

\def\sp{\>\>}


\def\half{{1 \over 2}}


\newdimen\pmboffset
\pmboffset 0.022em
\def\oldpmb#1{\setbox0=\hbox{#1}%
 \copy0\kern-\wd0 \kern\pmboffset\raise
 1.732\pmboffset\copy0\kern-\wd0 \kern\pmboffset\box0}


%
%
\def\appendix#1#2{\global\meqno=1\global\subsecno=0\xdef\secsym{\hbox{#1.}}
\bigbreak\bigskip\noindent{\bf Appendix. #2}\message{(#1. #2)}
\writetoca{Appendix {#1.} {#2}}\par\nobreak\medskip\nobreak}


\nref\BraatenYuanI{E. Braaten and T.C. Yuan, Phys. Rev. Lett. {\bf 71} (1993)
 1673.}
\nref\BCY{E. Braaten, K. Cheung and T.C. Yuan, Phys. Rev. {\bf D48} (1993)
 4230.}
\nref\BraatenYuanII{E. Braaten and T.C. Yuan, Phys. Rev. {\bf D50} (1994)
3295.}
\nref\Chen{Y.-Q. Chen, Phys. Rev. {\bf D48} (1993) 5181.}
\nref\Yuan{T.C. Yuan, UCD-94-2 (1994) unpublished.}
\nref\Braaten{E. Braaten, M. A. Doncheski, S. Fleming and M. L. Mangano,
 Fermilab-pub-94/135-T (1994), unpublished.}
\nref\Cacciari{M. Cacciari and M. Greco, FNT/T-94/13 (1994), unpublished.}
\nref\Roy{D.P. Roy and K. Sridhar, CERN-TH.7329/94 (1994),
 unpublished.}
\nref\Eichten{E. Eichten and F. Feinberg, Phys. Rev. {\bf D23} (1981) 2724.}
\nref\Buchmuller{W. Buchm\"uller, Phys. Lett. {\bf B112} (1982) 479.}
\nref\Isgur{S. Godfrey and N. Isgur, Phys. Rev. {\bf D32} (1985) 189.}
\nref\Kwong{W. Kwong, J.L. Rosner and C. Quigg, Ann. Rev. Nucl. and Part.
 Sci., {\bf 37} (1987) 325.}
\nref\Novikov{V.A. Novikov, L.B. Okun, M.A. Shifman, A.I. Vainshtein, M.B.
 Voloshin and V.I. Zakharov, Phys. Rept. {\bf 41C} (1978) 1.}
\nref\Cornell{E. Eichten, K. Gottfried, T. Kinoshita, K.D. Lane and T.-M.
 Yan, Phys. Rev. {\bf D17} (1978) 48.}
\nref\Bodwin{G.T. Bodwin, E. Braaten and G.P. Lepage, ANL-HEP-PR-94-24 (1994)
 unpublished.}
\nref\Kuhn{J. H. K\"uhn, J. Kaplan and E. G. O. Safiani, Nucl. Phys. {\bf
 B157} (1979) 125\semi
 B. Guberina, J.H. K\"uhn, R.D. Peccei and R. R\"uckl, Nucl. Phys. {\bf B174}
 (1980) 317.}
\nref\Cho{P. Cho, S. Trivedi and M. Wise, CALT-68-1943 (1994), unpublished.}
\nref\Close{F.E. Close, RAL-94-093 (1994), unpublished.}
\nref\Kuang{Y.-P. Kuang, S.F. Tuan and T.-M. Yan, Phys. Rev. {\bf D37} (1988)
 1210.}
\nref\PDB{Review of Particle Properties, Phys. Rev. {\bf D50}, Part I
 (1994).}


\nfig\singletgraphs{Lowest order color-singlet Feynman diagrams which mediate
$g \to \onedtwo$ fragmentation.}
\nfig\octetgraph{An example of a higher order Fock component contribution to
$g \to \onedtwo$ fragmentation.  The $\oneszero$ bound state which is formed
at short distances emits two soft gluons to become a physical $\onedtwo$
quarkonium.}
\nfig\gluefragplots{$g \to 1\onedtwo$ charmonium fragmentation functions
evaluated at (a) $\mu=M$ and (b) $\mu=10 M$.  The dotted, dot-dashed and
dashed curves represent the fragmentation functions for the $|h|=0$, 1 and
2 helicity components of the $J=2$ bound state, while the solid curve
illustrates their sum.}
\nfig\ptplot{Transverse momentum differential cross section for $1\onedtwo$
charmonium production at the Tevatron.  The dotted, dot-dashed and dashed
curves represent the cross sections for the $|h|=0$, 1 and 2 helicity
components of the $J=2$ bound state, while the solid curve illustrates their
sum.}


\def\CITTitle#1#2#3{\nopagenumbers\abstractfont
\hsize=\hstitle\rightline{#1}
\vskip 0.4in\centerline{\titlefont #2} \centerline{\titlefont #3}
\abstractfont\vskip .4in\pageno=0}

\CITTitle{{\baselineskip=12pt plus 1pt minus 1pt
  \vbox{\hbox{CALT-68-1954}\hbox{DOE RESEARCH AND}\hbox{DEVELOPMENT
  REPORT}}}}
{Gluon Fragmentation to $\onedtwo$ Quarkonia}{}
\centerline{
  Peter Cho\footnote{$^1$}{Work supported in part by a DuBridge Fellowship
  and by the U.S. Dept. of Energy under DOE Grant no.
  DE-FG03-92-ER40701.}
  and Mark B. Wise\footnote{$^2$}{Work supported in part by
  the U.S. Dept. of Energy under DOE Grant no. DE-FG03-92-ER40701.}}

\centerline{Lauritsen Laboratory}
\centerline{California Institute of Technology}
\centerline{Pasadena, CA  91125}

\vskip .2in
\centerline{\bf Abstract}
\bigskip

	Gluon fragmentation to heavy $J^{\P\C}=2^{-+}$ quarkonia is studied
herein.  We compute these D-wave states' polarized fragmentation functions and
find that they are enhanced by large numerical prefactors.  The prospects for
detecting the lowest lying $\onedtwo$ charmonium state at the Tevatron are
discussed.

\Date{10/94}

	One of the outstanding challenges in QCD is to understand the
process whereby colored quarks and gluons hadronize into colorless mesons and
baryons.  Until recently, the only basis for parton fragmentation intuition
came from simple models and empirical observations.  However within the
past two years, important progress has been made in understanding
hadronization from first principles \refs{\BraatenYuanI{--}\Yuan}.
It is now possible to calculate the fragmentation functions
which specify the probability for heavy quarks and gluons to hadronize into
quarkonium bound states starting from perturbative QCD.  These functions
involve nonperturbative matrix elements whose values must still be
extracted from experiment or the lattice.  But their dependence upon the
quarkonium longitudinal momentum fraction $z$ can be calculated to lowest order
in the strong interaction fine structure constant $\aS(\mQ)$ and the velocity
$v$ of the heavy constituent quark inside the bound state.  In principle,
higher order corrections associated with these two small expansion parameters
may be systematically evaluated as well. These developments have allowed a
range of hadronization issues to be explored in a limited but model
independent context.

	The first fragmentation functions to be computed from perturbative QCD
described the hadronization of gluons and heavy quarks into S-wave quarkonium
bound states \refs{\BraatenYuanI,\BCY}.  These $O(v^3)$ functions can be
used to predict the direct production of $\eta_c$ and $\J$ charmonia
as well as $\eta_b$ and $\Upsilon$ bottomonia at lepton and hadron colliders.
More recently, $O(v^5)$ P-wave fragmentation functions have also been
calculated \refs{\BraatenYuanII{--}\Yuan}.  In this paper, we extend the
ideas and methods developed in refs.~\refs{\BraatenYuanI{--}\Yuan} to the
D-wave sector.

	As $L=2$ fragmentation functions start at $O(v^7)$, they are generally
less important than those for $L=0$ and $L=1$ quarkonia.  P-wave contributions
to $Q \to \chi_\Q$ have been found to be quite suppressed compared to S-wave
terms \Yuan, and D-wave effects should be even smaller still.  On the other
hand, gluon fragmentation to $\chi_\Q$ is known to be phenomenologically
significant at hadron machines where the number of gluons in the initial
state is large.  Indeed, the dominant source of high $\pperp$ prompt $\J$'s
at the Tevatron is gluon fragmentation to $\chi_c$'s followed by
single photon emission \refs{\Braaten{--}\Roy}.  This prompt $\J$ mechanism
beats all others by almost two orders of magnitude.  It is therefore
interesting to examine the rate at which gluon fragmentation to D-wave
quarkonia occurs as well.

	We will focus in particular upon the production of the lowest lying
charmonium state with the quantum numbers $n=1$, $L=2$ and $S=0$.  This
$J^{\P\C}=2^{-+}$ meson has not yet been observed.  Quark model predictions
for its mass fall within the range $3.81 \GeV \le M \le 3.84 \GeV$ which
lies above the $D\Dbar$ threshold \refs{\Eichten{--}\Kwong}.  But since
its parity is odd, $1\onedtwo$ cannot decay to $D\Dbar$, for the two spinless
mesons would have to emerge in an even parity $L=2$ partial wave in order to
conserve angular momentum.  Moreover, the $n=1$ state lies below the
$D\Dbar\pi$ and $D{\Dbar}^*$ thresholds.  Therefore, $1\onedtwo$ charmonium
predominantly decays to lower $c\cbar$ levels or to light hadrons.  Its width
is consequently narrow.
\foot{Nonrelativistic potential models yield the estimate $\Gamma \simeq 0.34
\MeV$ for the total width of the $1\onedtwo$ charmonium state
\refs{\Novikov,\Cornell}.}

	The wavefunction for a physical $\onedtwo$ quarkonium can be
decomposed into a series of Fock state components:
\eqn\DwaveFockdecomp{|\onedtwo \sp {\rm Quarkonium} \rangle =
|(Q\Qbar)^{(1)}\rangle + O(v) |(Q\Qbar)^{(8)} g\rangle
+ O(v^2) |(Q\Qbar)^{(1,8)} g g\rangle + \cdots.}
The superscript labels on the $Q\Qbar$ pairs indicate whether the heavy quark
and antiquark reside within a color singlet or octet combination.  The
leading color singlet Fock component in \DwaveFockdecomp\ contributes to
$g \to \onedtwo$ fragmentation at $O(v^7)$ via the Feynman diagrams
illustrated in \singletgraphs.  These graphs mediate $\onedtwo$ production
through scattering processes like $ gg \to gg^* \to gg \, \onedtwo$.  Such
gluon fragmentation reactions dominate over lower order parton fusion
processes when the energy $q_0$ of the incoming off-shell gluon $g^*$ is
large, but its squared four-momentum $s=q^2$ is close to the bound state's
squared mass $M^2 \simeq (2\mQ)^2$.  In this kinematic regime, the total cross
section for $gg \to gg\onedtwo$ factorizes up to $O(s/q_0^2)$ corrections
which we neglect \BraatenYuanI:
\eqn\factor{\s(gg \to gg \onedtwo) = 2 \times \s(gg \to gg)
\int_0^1 dz D_{g \to \onedtwo}(z).}
The integrated fragmentation probability is thus simply given by the cross
section ratio
\eqn\xsectratio{\int_0^1 dz D_{g \to \onedtwo}(z) =
{\s(gg \to gg \onedtwo) \over 2\s(gg \to gg)}
= {1 \over 16 \pi^2} \int_{M^2}^\infty {ds \over s^2} \int_0^1 dz \,
\th\bigl(s - {M^2 \over z} \bigr) \sum |A(g^* \to \onedtwo g)|^2.}

	Higher $\onedtwo$ Fock state components also participate at $O(v^7)$
in $g \to \onedtwo$ fragmentation.  For example, the S-wave $(Q\Qbar)^{(1,8)}$
pair inside the $|(Q\Qbar)^{(1,8)} g g\rangle$ Fock state in
eqn.~\DwaveFockdecomp\ can be formed through the hard process
$g^* \to \oneszero g$ as pictured in \octetgraph.  It subsequently converts to
a physical $\onedtwo$ quarkonium through the spin-preserving emission of two
soft gluons.  The short distance formation of the $\oneszero$ bound state
takes place at $O(v^3)$ while the long distance gluon emissions each cost an
additional power of $v$ in the amplitude \Bodwin.  The graph in \octetgraph\
is therefore formally of the same order in the heavy quark velocity expansion
as the diagrams in \singletgraphs.

	Unfortunately, we do not know any rigorous way to determine the
nonperturbative matrix elements associated with the higher $\onedtwo$ Fock
state components.  However, previous experience with gluon and heavy quark
fragmentation to P-wave quarkonia provides some guidance.  In
$g \to \chi_\Q$ fragmentation, the $O(v^5)$ contribution from the color-octet
term in the $\chi_\Q$ wavefunction
\eqn\PwaveFockdecomp{|^3P_{0,1,2} \sp {\rm Quarkonium} \rangle =
|(Q\Qbar)^{(1)}\rangle + O(v) |(Q\Qbar)^{(8)} g\rangle + \cdots}
contains one less power of the short distance fine structure constant
$\aS(\mQ)$ than the $O(v^5)$ color-singlet term and is numerically more
important \BraatenYuanII.  On the other hand, the color-singlet term is much
larger than its color-octet counterpart in $Q \to \chi_\Q$ fragmentation where
both are $O\bigl(\aS(\mQ)^2\bigr)$ \Yuan.  Since the D-wave diagrams in
figs.~1 and 2 are also of the same order in $\aS(\mQ)$, the P-wave examples
suggest it is reasonable to assume that the latter graph is numerically small.
We will therefore simply neglect it in our analysis.

	We proceed to evaluate the two diagrams displayed in \singletgraphs\
by extending the Feynman rules for S and P-wave quarkonium processes derived
in ref.~\Kuhn\ to D-wave bound states.  Their sum yields the manifestly gauge
invariant amplitude
\eqn\glueamp{\eqalign{iA\bigl( g^*_a(q) \to \onedtwo(p) + g_b(p') \bigr) &=
-8 \sqrt{15 \over 2 \pi \Nc M} {g_s^2 \d_{ab} \over (s-M^2)^3} R''_2(0)
\varepsilon_\u(q) \varepsilon_\v(p')^* \varepsilon_{\a\b}^{(h)}(p)^* \cr
& \qquad \times \e^{\u\v\s\t} q_\s {p'}_\t (q+p')^\a (q+p')^\b. \cr}}
Here $\Nc=3$ denotes the number of colors, $g_s$ represents the strong
interaction coupling and $R''_2(0)$ equals the second derivative of the bound
state's radial wavefunction evaluated at the origin.
\foot{The $\onedtwo$'s polarization tensor is related to its nonrelativistic
wavefunction by
\eqn\poltensor{\int{d^3 \ell \over (2\pi)^3} \ell_\a \ell_\b
\psi_{2h}(\ell;p) = \sqrt{15 \over 8 \pi} \varepsilon^{(h)}_{\a\b}(p)
R''_2(0)}
where $\ell_\a$ represents the relative four-momentum between the heavy quark
and antiquark inside the bound state.  The polarization tensor's label
$h$ ranges over the helicity levels of the $J=2$ meson.}
Summing over the final gluon's color and polarization, we obtain the squared
amplitude
\eqn\sqrdamp{\eqalign{
{1 \over 16\pi^2} \sum |A\bigl( g^*(q) \to \onedtwo(p) + g(p') \bigr) |^2
&= -{1920 \aS^2 \over \pi\Nc} |R''_2(0)|^2
{(s-M^2)^2 g^{\u\v} + 4 s p^\u p^\v \over M(s-M^2)^6} \cr
&  \quad \times q^\a q^\b q^\c q^\d \sum \varepsilon_\u(q) \varepsilon_\v(q)^*
\sum \varepsilon^{(h)}_{\a\b}(p) \varepsilon^{(h)}_{\c\d}(p)^*. \cr}}
The remaining spin sums for the individual helicity levels of the
$J=2$ quarkonium may be evaluated using the covariant expressions given in
ref.~\Cho.  To the order at which we are working, we can also set
$p^\u = z q^\u + p_\perp$ and substitute
$p^\u p^\v \to -\half (1-z)(z s - M^2) g^{\u\v}$ \BraatenYuanI.  Then after
removing the factor $-g^{\u\v} \sum \varepsilon_\u(q) \varepsilon_\v(q)^*$ from
eqn.~\sqrdamp, inserting the squared amplitude into eqn.~\xsectratio,
and swapping the order of the $s$ and $z$ integrations, we simply read off
the polarized gluon fragmentation functions
\eqna\gluefragfuncs
$$ \eqalignno{
 D_{g \to \onedtwo^{(h=0)}}(z,M) &=
 {16 \aS(M)^2 \over \pi\Nc} {|R''_2(0)|^2 \over M^7} \Bigl[
 1080 z^{-3} - 2340 z^{-2} + 1680 z^{-1} - 450 + 37 z - 2 z^2 \cr
 & \quad + \bigl( 1080 z^{-4} - 2880 z^{-3} + 2760 z^{-2} - 1140 z^{-1}
 + 190 - 10 z \bigr) \log(1-z) \Bigr] & \cr
D_{g \to \onedtwo^{(|h|=1)}}(z,M) &=
 {64 \aS(M)^2 \over \pi\Nc} {|R''_2(0)|^2 \over M^7} (1-z) \Bigl[
 -360 z^{-3} + 420 z^{-2} - 120 z^{-1} + 5 + z \cr
 & \quad + (-360 z^{-4} + 600 z^{-3} - 300 z^{-2} + 45 z^{-1} \bigr)
 \log(1-z) \Bigr] & \cr
D_{g \to \onedtwo^{(|h|=2)}}(z,M) &=
 {32 \aS(M)^2 \over \pi\Nc} {|R''_2(0)|^2 \over M^7} (1-z) \Bigl[
 180 z^{-3} - 210 z^{-2} + 30 z^{-1} + 5 + 2 z \cr
 & \quad + \bigl( 180 z^{-4} - 300 z^{-3} + 120 z^{-2} \bigr) \log(1-z)
 \Bigr] & \gluefragfuncs a \cr} $$
and their unpolarized sum
$$ \eqalignno{D_{g \to \onedtwo}(z,M) &= {80 \aS(M)^2 \over \pi\Nc}
{|R''_2(0)|^2 \over M^7} \bigl[ 3z - 2z^2 + 2 (1-z) \log(1-z) \bigr]. &
\gluefragfuncs b} $$
The $z$ dependence of this last expression is precisely the same as that of
the $g \to \eta_c$ fragmentation function found in ref.~\BraatenYuanI.

	The functions in eqns.~\gluefragfuncs{a}\ and \gluefragfuncs{b}\ are
evaluated at the renormalization scale $\mu=M$ which corresponds to the minimum
allowed value for the fragmenting gluon's $\sqrt{s}$.  They may be evolved to
higher energies using the Altarelli-Parisi equation
\eqn\APeqn{\mu {d D_{g \to \onedtwo} \over d\mu}(z,\mu) = {\aS(\mu) \over
2\pi} \int_z^1 {dy \over y} P_{gg}(y) D_{g \to \onedtwo} \bigl({z \over
y},\mu \bigr)}
where
\eqn\split{P_{gg}(y) = 6 \Bigl[ {y \over (1-y)_+} + {1-y \over y} + y(1-y) +
{33 -2\nf \over 36} \d(1-y) \Bigr]}
denotes the gluon splitting function for $\nf$ active quark flavors.  The
$|h|=0,1$ and 2 polarized fragmentation functions for the lowest lying $n=1$
$\onedtwo$ charmonium state as well as their unpolarized sum are plotted
in figs.~3a and 3b for $\mu=M$ and $\mu=10 M$ respectively.  The
results displayed in the figure are based upon the parameter values
$M=3.82 \GeV$, $\aS(M)=0.256$ and $|R''_2(0)|^2=0.07 \GeV^7$ \Novikov.
Comparing the low and high energy curves, we see that the weights of all the
fragmentation functions are shifted toward lower values of $z$ as the
renormalization scale increases.  This behavior is consistent with the general
effect of Altarelli-Parisi running upon any fragmentation function.

	An approximate estimate for the rate of prompt $\onedtwo$ quarkonia
production at hadron colliders is given by the product of the total cross
section for gluon production and the initial integrated fragmentation
probability
\eqn\onedtwoprob{D_{g \to \onedtwo}(M) = {80 \aS^2(M) \over 3 \pi \Nc}
{|R''_2(0)|^2 \over M^7}.}
In the particular case of $1\onedtwo$ charmonium, we find
$D_{g \to 1\onedtwo}(M) \simeq 1.0 \times 10^{-6}$.  It is instructive to
compare this D-wave fragmentation probability with the corresponding S-wave
result \BraatenYuanI\
\eqn\etacprob{D_{g \to \eta_c}(\Meta) = {\aS^2(\Meta) \over 3 \pi \Nc}
{|R_0(0)|^2 \over \Meta^3} \simeq 5.3 \times 10^{-5}}
where $\Meta = 2.98 \GeV$, $\aS(\Meta) = 0.282$ and $|R_0(0)|^2 = 0.5
\GeV^3$.  Although the rate for $g \to 1\onedtwo$ is formally suppressed by
four powers of $v$ compared to that for $g \to \eta_c$, the probability of
the former is enhanced by a numerical prefactor of 80 relative to the latter.
Gluon fragmentation to the D-wave bound state is therefore larger than one
might have initially anticipated.

	A more precise prediction for $\onedtwo$ production may be obtained
by folding together the Altarelli-Parisi evolved fragmentation functions and
the gluon cross section $d\sigma(p\pbar \to g + X)/d\pperp$ into the
combination
\eqn\onedtwoxsect{{d\sigma(p\pbar \to \onedtwo^{(h)}+X)\over d\pperp} =
\int_0^1 dz{d\sigma(p\pbar \to g({\pperp / z}) +X,\mu) \over d\pperp}
D_{g\to \onedtwo^{(h)}}(z,\mu).}
This transverse momentum distribution is displayed in \ptplot\ for
$1\onedtwo$ charmonium production in the pseudorapidity range $|\eta|
\le 0.6$ at the Tevatron.  We have used the MRSD0 parton distribution function
evaluated at $\mu=M_\perp/z=\sqrt{M^2 + \pperp^2}/z$ to generate the
differential cross section shown in the figure.

	Since the gluon cross section is a steeply falling function of
$\pperp$, the main support for the integral in eqn.~\onedtwoxsect\ lies near
$z=1$.  Looking again at the fragmentation functions in figs.~3a and 3b,
we see that the $h=0$ helicity component dominates over the other helicity
levels for $z$ close to unity.  As a result, the $h=0$ differential cross
section in \ptplot\ is larger than its $|h|=1$ and $|h|=2$ counterparts.  Over
the transverse momentum interval $5 \GeV \le \pperp \le 30 \GeV$, the ratio
of these cross sections is approximately given by
\eqn\polratio{{d\s^{(h=0)} \over d\pperp} : {d\s^{(|h|=1)} \over d\pperp} :
{d\s^{(|h|=2)} \over d\pperp} \simeq 1.00 : 0.79 : 0.34.}
Recall that the helicity levels for an unpolarized $J=2$ state would be
populated according to
\eqn\unpolratio{{d\s^{(h=0)} \over d\pperp} : {d\s^{(|h|=1)} \over d\pperp} :
{d\s^{(|h|=2)} \over d\pperp} \simeq 1: 2: 2.}
Gluon fragmentation consequently induces a sizable $\onedtwo$ alignment.
This polarization can be observed in the angular distribution of photons
\eqn\angdist{{d\Gamma \over d\cos\th} \propto 1-0.33 \cos^2\th}
which result from the dominant E1 radiative transition $1\onedtwo \to
1 ^1P_1 + \gamma$ \Cho.  A measurement of this angular distribution would
provide a test of the $g \to \onedtwo$ fragmentation picture.

	The integral of $d\s(p\pbar \to 1\onedtwo+X)/d\pperp$ over the
transverse momentum range $\pperp \ge M$ where the results of our gluon
fragmentation calculation can be trusted yields 0.8 nb.  We should stress that
this integrated Tevatron cross section value represents a conservative lower
bound.  Higher Fock state contributions to $g \to 1\onedtwo$, charm quark
fragmentation and parton fusion processes will all enhance the production of
$1\onedtwo$ charmonia.  The D-wave state should therefore be produced
at a nonnegligible rate.

	Detecting $J^{\P\C}=2^{-+}$ charmonia will not be simple however.
One decay mode which might be observable is $1\onedtwo \to \psi' + \gamma$.
This M1 radiative transition connects the $\onedtwo$ initial state to the
$^3D_1$ component of the physical $\psi'$ \Close.  It is suppressed by the
small mixing angle which accompanies the $L=2$ component.  Another possible
mode which may be experimentally feasible to reconstruct is the following:
$$ \eqalign{1. & \cr 2. & \cr 3. & \cr} \quad
\eqalign{1\onedtwo & \to 1\onepone+ \g \cr
1 \onepone & \to \J + \pi^0 \cr
\J & \to \mu^+ \mu^- \cr}
\qquad\qquad
\eqalign{\Br &\simeq 0.80 \cr
\Br &\simeq 0.005 \cr
\Br &\simeq 0.06 \cr}
\sp
\eqalign{& [13,14] \cr
& [19] \cr
& [20]. \cr} $$
Approximate branching ratios for the steps in this decay chain are listed on
the right.  In order to overcome the small value for their product, a data
sample corresponding to a large integrated luminosity will have to be
collected.  The $1\onedtwo$ event rate may then be high enough to detect in
this channel.

	In conclusion, we have investigated gluon fragmentation to heavy
$\onedtwo$ quarkonia in perturbative QCD.  We have found that the polarized
$g \to \onedtwo$ fragmentation functions are enhanced by large numerical
prefactors and yield $J^{\P\C}=2^{-+}$ mesons which are significantly aligned.
These results provide motivation to search for these D-wave states in the
future.

\listfigs
\listrefs
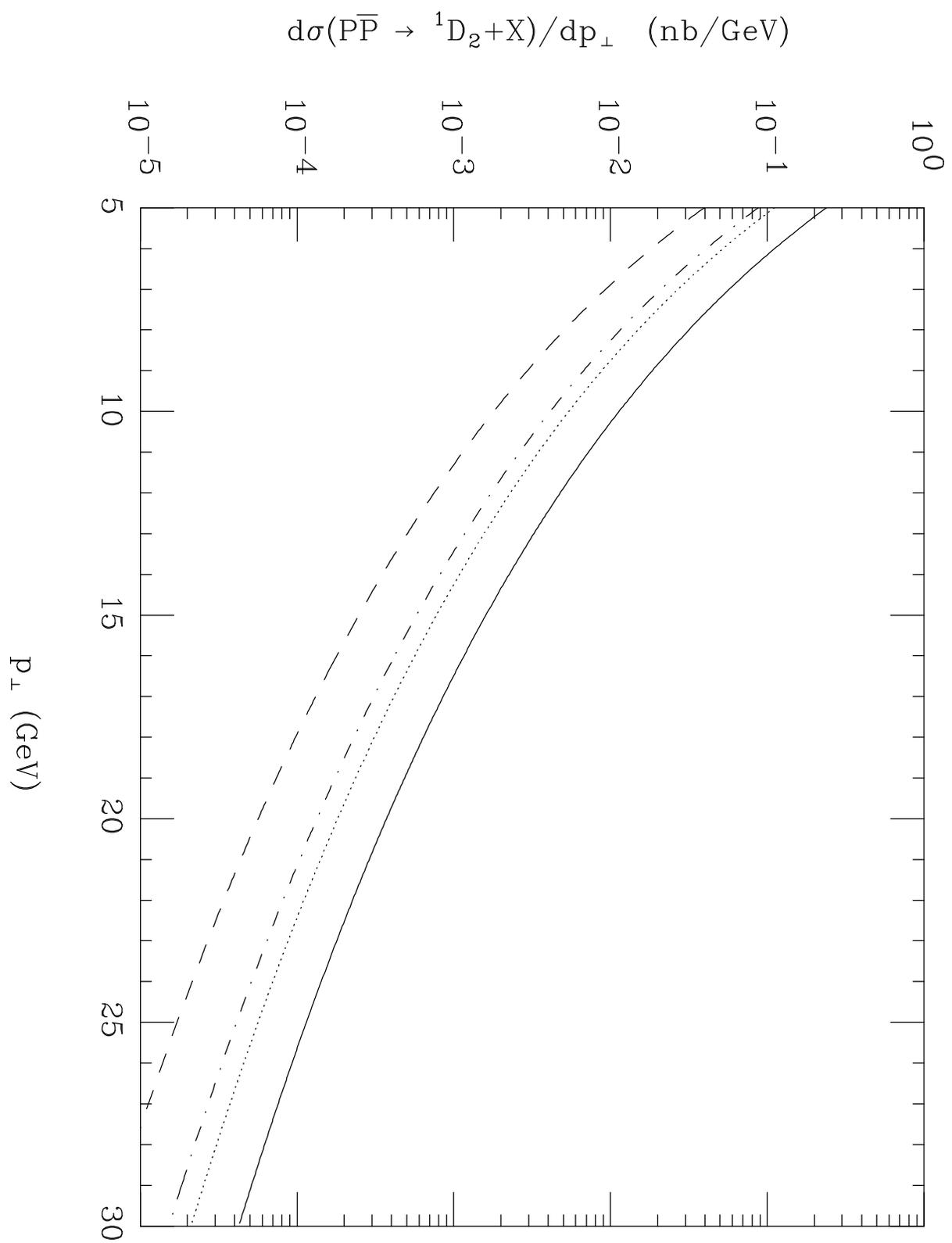
\bye